# Ab initio identification of putative human transcription factor binding sites by comparative genomics

D. Corà[1], C. Herrmann[2], C. Dieterich[3], F. Di Cunto[4], P. Provero[4] and M. Caselle[1§]

[1]Dipartimento di Fisica Teorica dell'Università degli Studi di Torino and INFN ,Via P. Giuria 1 – I 10125 Torino, Italy

[2]LGPD-IBDM, Université de la Méditerranée / CNRS, Campus de Luminy Case 907   - F-13288 Marseille Cedex 9,  France

[3]Max-Planck-Institute for Molecular Genetics,   Ihnestrasse 73 -  D-14195 Berlin, Germany

[4]Dipartimento di Genetica, Biologia e Biochimica dell'Università  di Torino ,  Via Santena 5 bis -  I-10126 Torino, Italy

[§]Corresponding author

Email addresses:

- DC: cora@to.infn.it
- CH: herrmann@ibdm.univ-mrs.fr
- CD: dieteric@molgen.mpg.de
- FDC: ferdinando.dicunto@unito.it
- PP: provero@to.infn.it
- MC: caselle@to.infn.it

# Abstract

**Background**

Understanding transcriptional regulation of gene expression is one of the greatest challenges of modern molecular biology. A central role in this mechanism is played by transcription factors, which typically bind to specific, short DNA sequence motifs usually located in the upstream region of the regulated genes. We discuss here a simple and powerful approach for the ab initio identification of these cis-regulatory motifs. The method we present integrates several elements: human-mouse comparison, statistical analysis of genomic sequences and the concept of coregulation. We apply it to a complete scan of the human genome.

**Results**

By using the catalogue of conserved upstream sequences collected in the CORG database we construct sets of genes sharing the same overrepresented motif (short DNA sequence) in their upstream regions both in human and in mouse. We perform this construction for all possible motifs from 5 to 8 nucleotides in length and then filter the resulting sets looking for two types of evidence of coregulation: first, we analyze the Gene Ontology annotation of the genes in the set, searching for statistically significant common annotations; second, we analyze the expression profiles of the genes in the set as measured by microarray experiments, searching for evidence of coexpression. The sets which pass one or both filters are conjectured to contain a significant fraction of coregulated genes, and the upstream motifs characterizing the sets are thus good candidates to be the binding sites of the TF's involved in such regulation.
In this way we find various known motifs and also some new candidate binding sites.

**Conclusions**

We have discussed a new integrated algorithm for the "ab initio" identification of transcription factor binding sites in the human genome. The method is based on three ingredients: comparative genomics, overrepresentation, different types of coregulation. The method is applied to a full-scan of the human genome, giving satisfactory results.

# Background

Understanding transcriptional regulation of gene expression is one of the greatest challenges of modern molecular biology. A central role in this mechanism is played by transcription factors (TF), which typically bind to specific, short DNA sequence motifs. These motifs are usually located in the upstream region of the regulated genes, although it is possible to find them also in the introns and in the 3' downstream region. They are often overrepresented, and appear in multiple copies inside the regulatory regions to form modules of cooperating items.

In these last years, the study of gene regulation has undergone a deep change of perspective [1,2]. While past studies usually dealt with individual regulatory interactions, it has become by now clear that the only way to understand the regulatory activity of the genome is to directly address the complex, combinatorial nature of the whole ensemble of TFs.

The identification of the cis-binding sequences and of the related TF's is a mandatory preliminary step toward this goal.

To this end it is becoming more and more important to construct tools able to
- address the problem on a genome wide scale
- keep under control the number of false positives to avoid an excessive increase of the noise to signal ratio
- use as input the statistical properties of the DNA sequences, thus avoiding, as far as possible, any other a priori assumption on the binding motifs.

However, the study cannot be based exclusively on the statistical features of the DNA regions presumably involved in transcriptional regulation, but must be complemented with independent information about gene regulation. In this respect three important sources of information may be used: the functional annotations collected in public databases, gene expression data on a global scale, and the so called 'phylogenetic footprinting' [3].

In fact large functional annotation databases and large-scale expression data provide a wealth of information about coregulation. This is a crucial point, since coregulated genes are likely to share similar transcriptional regulatory mechanisms. At the same time in these last years a growing interest has been attracted by the 'phylogenetic footprinting', i.e. the idea that functional sequences are preferentially conserved over the course of evolution by selective pressure. Comparison of orthologous gene sequences has been for a long time a standard tool in genomic analysis. Recently this comparative approach has been extended also to non coding regions, thanks to the progress of the sequencing programs. It is by now accepted that these non-coding conserved regions have an important regulatory role [4-9].

Several computational method have been proposed in the last few years to identify TF binding sites. These can be classified into two separate groups: enumerative methods, including the one we will present in this paper, explore all possible motifs up to a certain length (see for example [10-17]). The other large group consists of local search algorithms, including expectation maximization and various flavours of Gibbs sampling (see e.g. [18-21]).

We discuss here a simple and powerful approach for the identification of cis-regulatory motifs which fulfils the above requirements. It can be tuned to keep the number of false positives under control, and it allows us to study the transcriptional regulation of more than 10,000 genes of the human genome (which is a good approximation of a genome wide scale). The method is based on an "ab initio" study of the statistical properties of the regulatory sequences of the genes of interest. Together with the discussion of the method itself, we apply it to a full-genome analysis of the human case.

In particular in this paper, as a first step, we concentrated on the upstream sequences of the human genome (to be defined more precisely below), while we plan to extend these same tools to the downstream and intron regions in the future.

## Results and Discussion

Our proposed approach is based on three main ingredients:
 (1) human-mouse genomic comparison
 (2) statistical analysis of "motifs" (short DNA sequences) that are overrepresented in evolutionarily conserved regions upstream of orthologous genes
 (3) two complementary "filters" to infer coregulation: the distribution of Gene Ontology annotation terms and the results of a set of microarray experiments

The approach based on steps (2) and (3) above was successfully applied to the search for regulatory binding sites in yeast [14,15]. The human-mouse genetic comparison is crucial in extending the method to higher eukaryotes, since it is expected to greatly improve the signal/noise ratio by selecting for analysis those portions of the upstream regions that are more likely to be functionally relevant. Other algorithms taking advantage of phylogenetic footprinting to detect transcription factor binding sites have been published in [22,23].

As a final result of our analysis we obtain a set of motifs which survive one or both of the above filters, which we consider as our candidate binding sequences. The final step is then to cluster together these words to obtain consensus binding site sequences. This step allows, at the end of the whole process, to recover the intrinsic variability or regulatory motifs, that we know to be one of the most important features of binding sequences in higher eukaryotes.

A flow-chart of our algorithm is depicted in Figure 1.

The total number of motifs analyzed, after identifying each motif with its reverse complement and disregarding the self-overlapping ones, was 40,484. With the false discovery rate set at 10%, 373 of these turned out to be significant with at least one choice of scoring matrix (PAM1 or PAM10) and filter (Gene Ontology or microarray). 105 different Gene Ontology terms were involved, and 57 microarray time-points.

All the significant associations between motifs and Gene Ontology terms and microarray time points, for both scoring systems, are reported in Supplementary

Table 1 [see Additional file suptab1.txt]. The total number of such associations is 800, meaning that each motif was found, on average, about twice. Even if the various filters and scoring methods cannot be considered independent of each other, this fact is certainly encouraging in terms of the robustness of the method. These results are summarized in Tab. 1.

While the high degree of superposition between the results found with different choices is a clear indication of the robustness of our approach, the lists obtained are still significantly different: therefore the use of different filters/scoring matrices is useful in expanding the range of regulatory interactions explored by the algorithm.

In the supplementary table 2 we also provide the lists of the genes included in the sets found significant by one or more filter [see Additional file suptab2.txt].

Not surprisingly, in many instances several motifs, often very similar to each other, are associated to the same GO term or microarray timepoint. For all Gene Ontology terms and microarray point associated to one or more motifs, we constructed, when possible, a consensus binding sequence from the motifs associated to the term as explained in the Materials and Methods section. The results are presented in Tab. 2-4, for the three branches of the Gene Ontology and in Tab.5 for microarray time-points. For the latter, the consensus obtained is reported only when its length is at least 4. In many cases, in fact, the large number of motifs significantly associated to a microarray time-point causes the clustering algorithm to produce very short and rather uninformative consensus sequences. These results were produced by considering the PAM1 and PAM10 results together.

If our method is really able to identify genuine transcription factor binding sites, we would expect to find, among the surviving sets, at least some of the TF binding sites that are known to regulate the transcription of target genes through multiple occurrences in their promoters.

We focus here on some major examples.

E2Fs are transcription factors well known for their ability to regulate DNA replication by binding multiple sites in the promoters of the target genes [24]. Since the most abundant subpopulation of sets surviving the GO filter display a strong overrepresentation of DNA replication-related terms, it would be reasonable to expect that many of them are E2F binding sites. This was indeed the case, as the motifs TTGGCGC associated to many significant sets perfectly matched experimentally determined E2F binding sites as well as the consensus sequence found in the TRANSFAC[25] database. Significantly, some of these words were identified not only by the GO filter, but also by the microarray filtering scheme, confirming that our method is very robust in identifying the binding sites of this particular transcription factor.

It is interesting to see whether these motifs are found in the conserved parts of the upstream regions of experimentally verified targets of regulation by E2F. From the TRANSFAC database we identified 8 such targets, 6 of which are included in the PAM1 version of the CORG database: CAV1, MYC, DHFR, E2F2, RBL1 and CDC6. In five cases at least one of the motifs we find matching the E2F consensus can be found in the conserved part of the upstream regions, and in four cases many instances of the motifs are found: we find a total of 33 occurrences in the upstream region of MYC, 14 for E2F2, 11 for RBL1 and 11 for CDC6. Only one occurrence is found for DHFR, and none for CAV1. Similar results are found using the PAM10 version of CORG.

By performing these analyses, we observed that motifs characterized by the annotations 'chromatin' and 'nucleosome assembly', although obviously related to DNA replication, could not be reconciled with E2F binding sites, but included instead the motif AGAGCCTT and several similar ones. Since most of the annotated genes in the sets encoded for histone proteins, we speculated that these consensus could be part of a critical control element involved in the production of histones during DNA replication. One of the best known such elements is an evolutionary conserved inverted repeat found in the 3' untranslated region of histone mRNAs, controlling their stability during the cell cycle [26]. Surprisingly, our consensus sequence matched this element, raising the problem of how a 3' located regulatory element could be identified by our method. The reason is that histone genes form tight clusters in different chromosomal locations, and the distance between the initiator codon is in many cases below the 15000 bp limit used by our algorithm. Although of serendipitous nature, this result underscores two important points. The first is that our method is able to identify not only regular transcription factor binding sites, but also other less conventional regulatory elements characterized by a motif repetition. The second is that our approach could be systematically extended to other gene regions, such as the 3' untranslated and the introns.

The highly heterogeneous annotation associated to the sets surviving the GO filter strongly suggests that our method can potentially identify relevant binding sites for known and/or unknown transcription factors in the promoter of groups of genes involved in a wide variety of biological processes, such as tissue and organelle-specific transcription.

For instance, we identified many sets significantly enriched for genes involved in muscle development and/or functions (see Tab. 6). Interestingly, one of them (AGCAGG, associated to the term "sarcomere") is compatible with the binding site of the well known muscular master genes MyoD and Myf5 [27] as represented by a mixture of the TRANSFAC matrices M00184 and M0001, while the others had no significant match in the TRANSFAC database.

Another example are the different motifs associated with the annotations "endoplasmic reticulum", "protein transport" and "intracellular protein transport"(Tab. 7). Three of them (ACGTG, CCACGTCA and GACGTGGC) with known binding sites of ATF6 (TRANSFAC matrix M00483), a strongly conserved transcription factor involved in endoplasmic reticulum function [28]. The others don't show significant overlapping with TRANSFAC, suggesting that they are new putative cis elements important for regulating ER genes.

It is important to notice that in some instances, even though no hypothesis on the precise transcription factor can be formulated, it is at least possible to conjecture the general structural class to which the TF belongs. For example, the word GGGGGGGT, associated with the annotation "organogenesis" , is consistent with the binding sites of many zinc finger transcription factors, such as Zic1, Zic3 and MZF1 [29], thus suggesting that some of the genes in the set are transcriptional target for a member of this particular family of transcription factors.

It is interesting to investigate the distribution of the distance of the motifs identified by our algorithm from the TSS of the corresponding gene. For all motifs found significant and for all genes in which the motif is overrepresented we computed the distance between the locations in which the motif is found and the TSS of the gene. All these data are represented as a histogram in Fig. 2. The motifs are very obviously concentrated near the TSS. This fact suggests that the choice to cut at 15,000 bp the length of the upstream regions considered is unlikely to decrease the signal significantly. The data shown are for PAM1, but the ones for PAM10 do not differ in any significant way.

Taken together, these results suggest that our approach has the potential to identify new critical regulatory elements for genes involved in a wide variety of biological processes.

## Conclusions

We have discussed a new algorithm for the "ab initio" identification of transcription factor binding sites in the human genome. The method is based on three ingredients:

- the so called phylogenetic footprinting, i.e. the idea that functional sequences are preferentially conserved over the course of evolution by selective pressure.

- the overrepresentation criterion, i.e. the observation that binding sequences are usually overrepresented in the upstream region of the genes that are regulated by the corresponding transcription factors.

- the coregulation test, i.e. the use of coregulation (detected by using GO categories or microarray data) as a criterion to select the true positive binding sequences

Experience with yeast [14,15] suggests that our method is characterized by a low rate of false positives but, presumably, a rather high number of false negatives. The reason for this is that the basic ingredients of our analysis are motifs defined as completely specified sequences. This requires the motifs to be overrepresented, in the upstream region of a gene in order to be selected for our analysis and thus limits our candidates to a subset of all possible motifs. Our method is therefore complementary with respect to the standard approaches to binding sequences identification which use weighted matrices instead of completely specified motifs, since these typically have problems in detecting the true positive signals from the statistical background noise.

The variability of the motifs, which is a fundamental feature of Eukariotic binding sequences and is neglected at the beginning of our algorithm is recovered at the end thanks to the careful consensus reconstruction discussed in the previous section. These are the major novelties of our approach.

We consider as an encouraging validation test of our procedure the fact that several known TF binding sequences are found with our method. This makes us confident about the reliability of the other candidates that we found. Needless to say, these should be validated with suitable experimental tests. Indeed we think our "ab initio" approach could be of value as a preliminary test for any experimental search of binding sequences.

Several improvements of the present algorithm are possible. In particular it would be interesting to extend our analysis to other regions besides the 5' upstream one (the results on the control element of histones discussed above clearly indicate that this would be a fruitful research direction). In this respect the most natural candidates are the 3' downstream regions and the first intronic interval.

The method could be extended without major modifications to motifs with gaps, as considered in [40]. Extension to longer motifs would also be important: however the extension of the algorithm to motifs significantly longer than the ones considered here should probably take into account motif variability from the start, which would in turn imply a significant increase in computational complexity. We are currently investigating some possible ways of overcoming this problem.

Similarly, it would be important to address the combinatorial nature of transcriptional regulation by studying the correlation of overrepresented words along the lines discussed for instance in [30,31]. It is only by looking at the intricated network of interactions as a whole that one can hope to understand the collective behaviours leading to the tight and impressively efficient regulation of gene expression in higher eukaryotes and in particular in mammalians. It also clear that the algorithm can in principle be applied to any pair of closely related organisms.

We plan to address these issues in future work.

## Methods

**Construction of the new release of the CORG database**

Definition of upstream regions and conserved non-coding blocks:

An upstream region is a sequence window that contains 5' genomic DNA extending from the start of translation of each individual transcript. The maximal size of an upstream region is taken to be 15kbp. This upper bound stems from the observation that most promoter regions are less than 15,000 bp away from the start of translation [32]. Evidently, upstream regions may be smaller since they are bounded by the size of the intergenic region under consideration. Given this definition, upstream regions of different transcripts of the same gene or transcripts belonging to neighbouring genes could overlap. This is taken into account when compiling the conservation information by cutting the upstream region short of 15 kbp when necessary. All man and mouse DNA sequences were retrieved from the NCBI genome assemblies (NCBI33 and mNCBI30). Gene annotations were obtained from the EnsEMBL databases (release 17).

Orthologous man/mouse upstream regions were scanned for significant local similarities. We prefer a local alignment approach over a global one. That is we do not constrain the arrangement of putative regulatory modules. We denote these similarities as Conserved Non-coding sequence Blocks (CNBs). CNBs are computed with an implementation of the algorithm of Waterman and Eggert [33], which extends the well known Smith-Waterman algorithm to suboptimal alignments. The two scoring matrices used in the computation are derived from the Kimura two-parameter model and are normalized to a distance of 1 PAM and 10 PAM, respectively. The two matrices yield alignments of differing stringency with an expected level of identity of 99% for 1 PAM vs. 90.7 % for 10 PAM. Gap penalties were set to 11x match score for opening a gap and 0.1x match score for extending one. On average 8% (1 PAM) vs. 18% (10 PAM) of each upstream region (excluding repeats) is covered with CNBs.

An assessment of statistical significance of alignment scores was introduced to discriminate "true" from random alignments. Waterman and Vingron [34] showed that scores of local suboptimal alignments follow approximately the order statistics of a Poisson distribution. This facilitates the calculation of p-values by simulating random scores. We applied a P-value cutoff of 0.001.

Further details on the derivation of the data set can be found in [35]. Most of the data are part of the CORG database and can be accessed via the website (http://corg.molgen.mpg.de).

**Construction of the sets**

The first step in the algorithm is the construction of sets of genes associated to all possible motifs. In this work a motif is defined as a short (5-8 bps), completely specified DNA sequence. The set associated to the motif *m* consists of all genes such that *m* is overrepresented, in the sense defined below, in the CNBs upstream of the genes. Motifs are always read on both strands, and therefore the sets associated to a motif and to its reverse complement coincide by definition. All genes for which one ore more CNBs were available were examined and assigned to one or more sets: 11,265 genes are included in the PAM1 version of CORG, and 13,294 in the PAM10 one.

The CORG database includes many rather long entries, up to several hundred bps. It is likely that many of these are actually exons. Since the inclusion of long exons would decrease our signal/noise ratio, we discarded all entries of length greater than 200 bps. We also eliminated multiple overlapping entries so that, as a final result of this preliminary step, each nucleotide in each conserved upstream region has exactly the same statistical weight. With this choice we end up with a total of 389560 distinct CNBs in the PAM10 case out of which 9155 (2.3%) have a length greater than 200bp and, according to the strategy discussed above, were discarded in the following steps of our analysis. In the PAM1 case we find a total of 203417 CNBs out of which 3408 (1.7%) have a length greater than 200b. As expected the proportion of CNBs larger than 200bp decreases as the stringency of the alignments increases.

The definition of overrepresentation of a motif is the same that we used in Ref. [14] and [15], and was originally introduced in Ref.[16]. It is based on the frequency *f(m)* of the motif in all the CNBs contained in the database. For each gene we count the occurrences of m in the CNBs associated to the gene; then we compute the probability *P* of finding as many or more occurrences, based on a binomial distribution. The parameters of the binomial distribution are chosen as follows: *f(m)* is the success probability at each trial, and the number of trials is equal to the number of motifs that can be read in the CNBs associated to the gene. The use of the binomial approximation is based on the assumption of independence between successive trials. While rigorously speaking such assumption is never correct, it leads to serious errors only for periodic motifs, that are likely to be repeated several times in a row on the sequence. Therefore we did not include in our analysis the motifs that can be found repeated (possibly as their reverse complement) at a distance of 1, 2 or 3 bps (for example, respectively, CCCCC, ACGTA, CATCA).

If $P < 0.01$ we include the gene in the set labeled by the motif *m*. Notice that no biological significance is ascribed to these sets before they are selected for evidence of coregulation as explained below: therefore the choice of the cutoff on *P* can be arbitrarily lenient. Based on previous experience, we set the cutoff at $P = 0.01$. As it can be expected from the number of genes analyzed, essentially all possible motifs turn out to be overrepresented in some genes with this cutoff; however only a small fraction of them are selected by the GO and microarray filters and thus identified as candidate binding sites.

At this point we have thus obtained, for each possible motif *m*, a set of genes such that *m* is overrepresented in the CNBs of the genes in the set. The next step consists in looking for evidence of coregulation of the genes included in each set.

**The Gene Ontology filter**

As a first filter to select the sets whose genes are functionally related, and hence likely co-regulated, we analyze the prevalence of Gene Ontology (GO) annotations terms [36] in each set. For all GO terms associated to the genes of a set we perform an exact Fisher's test to determine whether the term appears in the set significantly more often than expected by chance. More precisely the Fisher's test gives us the probability *P* of obtaining an equal or greater number of genes annotated to the term in a set made of the same number of genes, but selected at random from the database. If *P* is statistically significant, then we can postulate the existence of a correlation between the overrepresentation of the motif m labeling the set and the functional characterization of the genes in the set, and hence include m in the list of candidate binding sites found by the algorithm.

Since this test is performed for all GO term and all sets, multiple testing is certainly an important issue. It is made rather non-trivial by the fact that the tests made on different GO terms are far from being independent of each other (think for example of testing the same set of genes for overrepresentation of the terms "cell cycle" and "DNA replication"). We chose to approach this issue with a safe, brute-force method based on random sampling, previously used in Ref. [15]: we generated a large sample of randomly selected gene sets of the typical size of our sets and we used it to estimate the number of false discoveries to be expected as a function of the cutoff on

P-values. This allowed us to tune the cutoff on the Fisher's test P-values to obtain the desired value of the FDR (False Discovery Rate), which for the results we present is 10%.

**The Microarray filter**

An alternative and complementary filter to select candidate binding sites of our method uses microarray data. The assumption is that the distribution of expression values of a set of co-regulated genes is significantly different from the distribution of expression values of the whole genome. We used microarray data from the Stanford human cell-cyle experiment [37], consisting of 114 microarrays. We used the labels available in the raw data file (Unigene identifier and HUGO symbol when available) to make the correspondence with the Ensembl clusters used to identify the genes in the sets.

For each set and each microarray experiment, the comparison between the distribution of the data for the set and for the wholes genome was performed using the non parametric Kolmogorov-Smirnov (KS) test on the distribution of log(R), where R is the red/green normalized ratio. The goal is to identify the sets (and thus the corresponding motifs) showing an expression pattern significantly different from the background distribution (i.e. from the whole genome expression pattern for that particular microarray experiment). The non parametric KS test is the best suited tool for this type of analysis since it makes it possible to compare the expression levels measured in a given experiment without any a-priori assumption about the distribution of the data. Moreover the KS test looks for significant differences in the whole distribution (the statistic used is the largest difference between the cumulative distributions): therefore, at least in principle, it is able to detect subtle differences which would not be detected by tests based, for example, simply on the average expression level. However, like most non-parametric tests, the KS test is generally less potent than parametric tests, and hence requires a very strong signal to turn out significant. In particular, it is more likely to be successful in detecting the differential expression of large sets of genes, like our own. The KS test on expression data was previously used to identify candidate binding sites in Ref. [17].

Finally we evaluated the False Discovery Rate (FDR) by using the standard Benjamini-Hochberg method [38], setting a FDR threshold of 10%.

**Construction of consensus sequences for the binding sites**

In many cases several words, similar to each other, are found to be significantly associated to the same Gene Ontology term, or to the same microarray experiment. In such cases it is natural to assemble such words into a consensus sequence for the candidate binding site. This was systematically done for each Gene Ontology term in the following way: all the words associated to the same Gene Ontology term were aligned using the *wconsensus* [39] package. We selected the wconsensus results in the following way: the best matrix found by wconsensus was accepted if its expected frequency was less than 0.001; other matrices were also accepted if they exceeded such significance and they were generated from motifs that did not enter the previously accepted matrices. Therefore the algorithm is in principle capable of generating more than one consensus from a group of motifs. However in practice this never happens in our case: either one or no consensus sequence was produced for each group of motifs.

The same approach gives less satisfactory results when applied to the words associated to the same microarray experiment, the reason being that several microarray experiments turn out to be associated to a large number of rather different motifs, which cannot produce a single, meaningful consensus sequence. This is hardly surprising based on the results of the same approach applied to yeast [14], where the analysis of a rather small set of microarray experiments revealed many unrelated binding sites. Indeed genes regulated by several different transcription factor can be expected to show differential expression in the same experimental conditions. For several time-points, the best consensus was a three-letter sequence of dubious informative value. Only for six time-points we obtained a consensus of length 4 or higher.

Additional material and raw data are available at:
*http://www.to.infn.it/ftbio/tf_human/supplementary.html*

## Acknowledgements

P.P. is a Lagrange Fellow of the I.S.I. foundation. We thanks Enrico Curiotto for helpful discussion.

# Figures

**Figure 1 - Flow-chart of the algorithm**

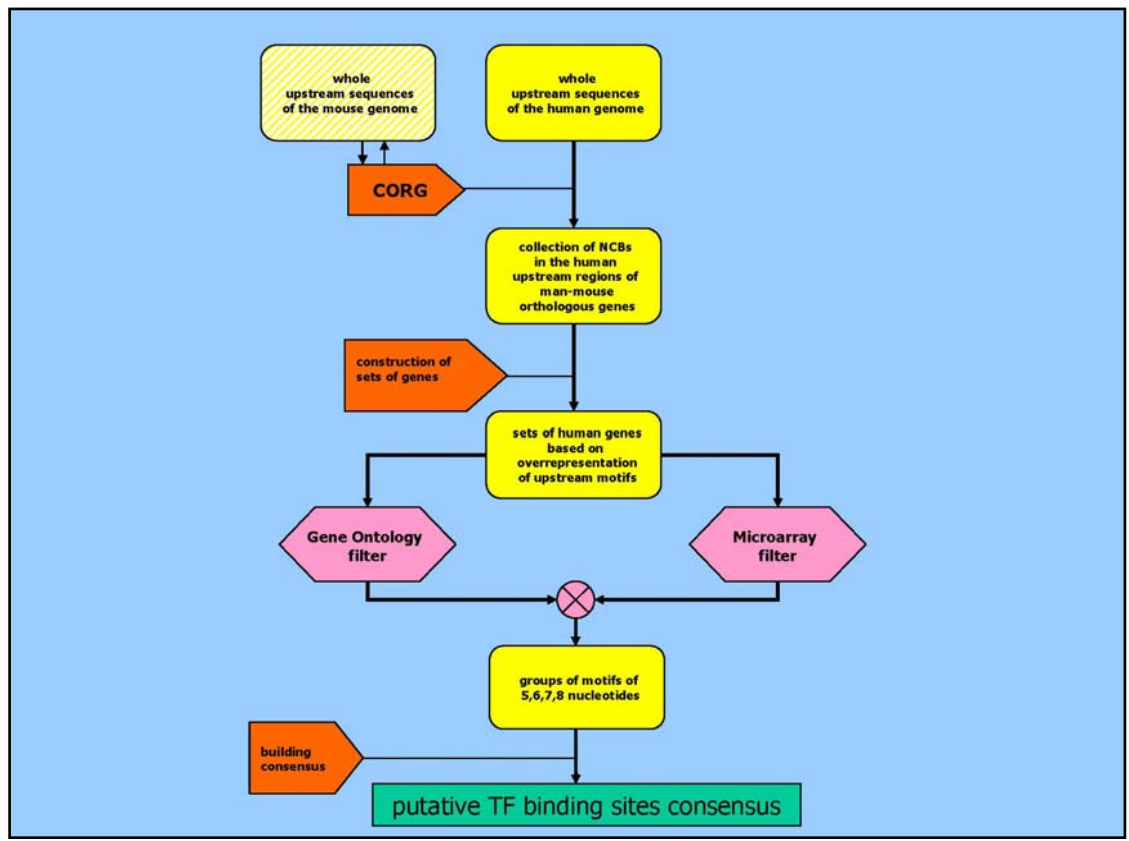

**Figure 2 - Histogram of the distance from the TSS of the motifs found significant by the algorithm (see text).**

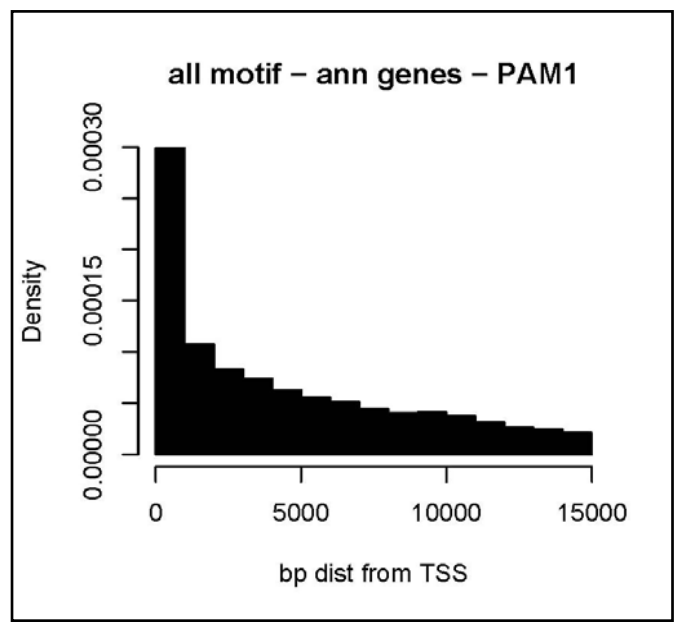

# Tables

## Table 1 - Number of significant motifs found with the four scoring matrix/filter combinations and their intersection

The third line contains the number of motif identified using both the PAM1 and PAM10 scoring system. The third column shows the number of motif identified by both the Gene Ontology and the microarray filter.

|              | GO  | MA  | GO & MA |
|--------------|-----|-----|---------|
| PAM1         | 139 | 61  | 29      |
| PAM10        | 93  | 181 | 55      |
| PAM1 & PAM10 | 42  | 38  | 17      |

## Table 2 - Consensus binding sites corresponding to GO terms in the *biological process* branch of the Gene Ontology

For each GO term we display either the consensus sequence obtained from wconsensus, or the longest motif associated to the term if the consensus sequence was not significant enough as defined in the text. The third column is the logarithm of the expected frequency of the alignment as given by wconsensus, if exists. The fourth column contains the number A of motifs which were used in the alignment and the total number B of motifs associated to the term in the format A/B. For this table the data obtained with PAM1 and PAM10 are considered together.

| actin filament-based process | GGGATTA | - | 1/1 |
|---|---|---|---|
| ATP metabolism | CCGTCCC | - | 1/1 |
| biosynthesis | CGCACG | - | 1/1 |
| cell growth and/or maintenance | CTTCA | - | 1/1 |
| cell motility | AGGGG | - | 1/2 |
| defense response | AGGAA | - | 1/1 |
| development | CCCC | -32,3389 | 16/16 |
| DNA metabolism | TTCCCGC | -35,3236 | 6/7 |
| DNA replication and chromosome cycle | TTCCCGCG | -17,6184 | 4/4 |
| DNA replication initiation | GCGCGAAA | - | 1/1 |
| enzyme linked receptor protein signaling pathway | AGGGGG | - | 1/1 |
| epidermal differentiation | AGGCA | - | 1/1 |
| frizzled-2 signaling pathway | GCTGGAGA | - | 1/1 |
| glycoprotein catabolism | CTGACCTA | - | 1/1 |
| heterophilic cell adhesion | CTAAACTC | - | 1/1 |
| immune response | GAAAC | - | 1/1 |
| intracellular protein transport | CCACGTC | -7,62462 | 2/2 |
| L-amino acid transport | ACTTTG | - | 1/1 |
| macromolecule catabolism | GACTC | - | 1/1 |
| metabolism | CGGAAG | - | 1/2 |
| metabolism | CGGGCCCG | - | 1/2 |
| mitotic cell cycle | TCCCGCCA | - | 1/1 |
| muscle development | CCAAG | - | 1/1 |
| negative regulation of cell growth | AACGACT | - | 1/1 |
| nucleobase\, nucleoside\, nucleotide and nucleic acid | AACGG | - | 1/4 |
| metabolism | | | |
| nucleosome assembly | GGCTCT | -92,8905 | 27/40 |
| organogenesis | ACCCCCCC | - | 1/2 |
| perception of chemical substance | TCTAA | - | 1/1 |
| phototransduction | AAGRGGCC | -12,0169 | 6/6 |
| pinocytosis | CTTACGA | -7,62462 | 2/2 |
| potassium ion transport | CCAAG | - | 1/1 |
| protein biosynthesis | CGGAAG | - | 1/1 |
| protein transport | CCCAG | - | 1/1 |
| regulation of apoptosis | CATAG | - | 1/1 |
| regulation of protein kinase activity | AAAAG | - | 1/1 |
| regulation of translation | CGTGCTTC | - | 1/1 |
| ribonucleotide metabolism | CTTGATCC | - | 1/1 |
| RNA localization | ACGCCG | - | 1/1 |
| synaptogenesis | AGCGCCAC | - | 1/1 |
| transcription | CCGAG | - | 1/1 |
| transcription\, DNA-dependent | CCGAG | - | 1/2 |
| translation | ACTTCCGG | - | 1/1 |
| two-component signal transduction system (phosphorelay) | CACACGGG | - | 1/1 |
| vision | AATCCCT | - | 1/1 |

**Table 3 - Consensus binding sites corresponding to GO terms in the *cellular component* branch of the Gene Ontology**

For each GO term we display either the consensus sequence obtained from wconsensus, or the longest motif associated to the term if the consensus sequence was not significant enough as defined in the text. The third column is the logarithm of the expected frequency of the alignment as given by wconsensus, if exists. The fourth column contains the number A of motifs which were used in the alignment and the total number B of motifs associated to the term in the format A/B. For this table the data obtained with PAM1 and PAM10 are considered together.

| | | | |
|---|---|---|---|
| actin cytoskeleton | AGGAC | - | 1/1 |
| chromatin | GGCTC | -9,99174 | 3/3 |
| chromosome | GGCGGAA | - | 1/2 |
| chromosome\, pericentric region | CAAATAGA | - | 1/1 |
| clathrin-coated vesicle | ATGGCA | - | 1/1 |
| collagen | GGACC | - | 1/1 |
| COPI-coated vesicle | CTCAGAG | - | 1/1 |
| cytosol | CGAAAGC | - | 1/2 |
| cytosolic large ribosomal subunit (sensu Eukarya) | CGGAGGAG | - | 1/3 |
| cytosolic ribosome (sensu Eukarya) | TTTCCG | -11,6127 | 4/5 |
| endoplasmic reticulum | GACGTGGC | - | 1/4 |
| eukaryotic 43S preinitiation complex | CGGAAAA | - | 1/2 |
| eukaryotic 48S initiation complex | GGGCGGAA | - | 1/1 |
| eukaryotic translation initiation factor 3 complex | CACCTCCG | - | 1/4 |
| external encapsulating structure | GTATCTA | - | 1/1 |
| extracellular matrix | CAAATG | - | 1/2 |
| extracellular space | GGGAA | - | 1/1 |

| fibrillar collagen | ACCCT | - | 1/1 |
| Golgi lumen | CAACAT | -8,99019 | 3/4 |
| heterogeneous nuclear ribonucleoprotein complex | AATGGCG | - | 1/4 |
| inner membrane | ACCGGCT | - | 1/1 |
| integral to membrane | ATCTCTG | - | 1/4 |
| integral to nuclear inner membrane | ACCTGAG | - | 1/2 |
| intracellular | CGGAAGCG | -23,1425 | 5/15 |
| lytic vacuole | GATTCA | - | 1/1 |
| membrane | CCTGGC | - | 1/6 |
| minor (U12-dependent) spliceosome complex | ATTGCG | - | 1/1 |
| mitochondrial inner membrane presequence translocase complex | ACGGGAA | - | 1/2 |
| mitochondrion | AAGTTGC | - | 1/2 |
| muscle fiber | CCTCAG | - | 1/1 |
| muscle myosin | CAGAG | - | 1/1 |
| muscle thin filament tropomyosin | TCCTCCA | - | 1/1 |
| nuclear chromatin | ATTGAG | - | 1/1 |
| nucleosome | GGCTCT | -85,6578 | 28/45 |
| nucleus | CACCAATC | - | 1/5 |
| plasma membrane | CTCCC | - | 1/1 |
| replisome | TCCCGCCA | - | 1/1 |
| ribonucleoprotein complex | CSGAA | -18,8768 | 6/8 |
| ribosome | CGTGTAG | - | 1/3 |
| sarcomere | AGCAGG | - | 1/2 |
| small ribosomal subunit | GGCGGAA | - | 1/2 |
| synaptic vesicle | ACCAGAAT | - | 1/1 |
| synaptonemal complex | GGTCTTA | - | 1/1 |
| vesicle coat | ACTGCCT | - | 1/1 |
| voltage-gated calcium channel complex | CCTCCC | - | 1/1 |

**Table 4 - Consensus binding sites corresponding to GO terms in the *molecular function* branch of the Gene Ontology**

For each GO term we display either the consensus sequence obtained from wconsensus, or the longest motif associated to the term if the consensus sequence was not significant enough as defined in the text. The third column is the logarithm of the expected frequency of the alignment as given by wconsensus, if exists. The fourth column contains the number A of motifs which were used in the alignment and the total number B of motifs associated to the term in the format A/B. For this table the data obtained with PAM1 and PAM10 are considered together.

| calcium-activated potassium channel activity | GCCACA | - | 1/1 |
| chemoattractant activity | GAATTTCC | - | 1/1 |
| G-protein coupled receptor activity | AATAG | - | 1/1 |
| ligand-dependent nuclear receptor activity | CAGGG | - | 1/1 |
| nucleic acid binding | CGGGAG | - | 1/2 |
| pancreatic ribonuclease activity | AACTACTC | - | 1/1 |
| phosphatidylinositol-4\,5-bisphosphate 3-kinase activity | AAGGA | - | 1/1 |
| retinoic acid receptor activity | ACCCA | - | 1/1 |

| RNA binding | ATGCG | - | 1/1 |
| serine-type endopeptidase activity | CAGAGGG | - | 1/1 |
| single-stranded DNA binding | AAACC | - | 1/1 |
| surfactant activity | ACTCACCC | - | 1/1 |
| translation factor activity\, nucleic acid binding | CGGAAG | - | 1/1 |
| uncoupling protein activity | GACGTAGC | - | 1/1 |

**Table 5 - Consensus binding sites corresponding to microarray time-points**

Only time-points for which the clustering algorithm produced a consensus sequence of length 4 or more are shown.

| Timepoint | consensus | sequences used | ln (expected freq.) |
|---|---|---|---|
| t = 23 | CTGG | 4/7 | -7,99646 |
| t = 50 | CCMCA | 5/15 | -9,71859 |
| t = 61 | SCCAGG | 12/43 | -18,6948 |
| t = 89 | CWGGG | 17/23 | -11,1386 |
| t = 100 | CCCWG | 12/31 | -12,5918 |
| t = 107 | CGGM | 13/14 | -14,7383 |

**Table 6 - Words associated to "muscle development" and related terms**

| AGCAGG | sarcomere |
|---|---|
| CCAAG | sarcomere |
| CCAAG | muscle development |
| TCCTCCA | muscle thin filament tropomyosin |

**Table 7 - Words associated to "endoplasmic reticulum", "protein transport" and "intracellular protein transport"**

| AAGTTGG | endoplasmic reticulum |
|---|---|
| AATCGGC | endoplasmic reticulum |
| ATCAGCG | endoplasmic reticulum |
| CGCAG | endoplasmic reticulum |
| GACGTGGC | endoplasmic reticulum |
| ACGTG | intracellular protein transport |
| CCACGTCA | intracellular protein transport |
| GACGTGGC | intracellular protein transport |
| CCCAG | protein transport |

## Additional files

**Additional file 1 – suptab1.txt**

The complete list of significant motif/GO term and motif/microarray time-point associations. The first column is the motif (to be considered coinciding with its reverse complement); the second column is the scoring matrix; the third is the type of filter ("GO" for Gene Ontology or "MA" for microarray); the fourth is the GO term or the microarray experiment; for GO terms, the fifth column contains the GO branch ("C": cellular component; "F": molecular function; "P": biological process); the sixth column is $-\log_{10}$ of the P-value of the test determining the significativit of the motif (Fisher's test for Gene Ontology, Kolmogorov-Smirnov for microarrays).

**Additional file 2 – suptab2.txt**

The complete list of sets corresponding to the significant motifs. Each gene in each significant set is represented by its EnsEmbl ID.